\documentclass[]{iscram}

\usepackage[utf8]{inputenc}

\usepackage{url}
\usepackage{lipsum}
\usepackage{tikz}
\usepackage{fancyvrb}
\usepackage{listings}
\usepackage{enumitem}
\usepackage{tcolorbox}
\usepackage{multicol}
\usepackage{siunitx}
\usepackage{caption}
\usepackage{subfigure}
\usepackage{todonotes}
\usepackage{graphicx}
\usepackage{xurl}

\lstdefinestyle{common}{
  xleftmargin=.5em,
  xrightmargin=.5em,
  frame=single,framesep=.5em,framerule=0pt,
  fancyvrb=true,
  basicstyle=\ttfamily,
  keywordstyle=\color{cyan!50!blue!75!black}\bfseries,
  commentstyle=\color{red!50!black}\itshape,
  stringstyle=\ttfamily\color{green!50!black},
  numbers=none,
  showspaces=false,
  showstringspaces=false,
  fontadjust=true,
  keepspaces=true,
  flexiblecolumns=true,
  emphstyle=\color{red},
}
\lstdefinestyle{TeX}{
  style=common,
  backgroundcolor=\color{blue!5},
  aboveskip=5pt,
  belowskip=5pt,
  language=[LaTeX]TeX,
  moretexcs={
    abstract, addbibresource, iscramset, keywords, mainmatter,
    maketitle, printbibliography, subsection, subsubsection, url,
    urldef, href, includegraphics, ldots, parencite, citeauthor,
    citeyear, citetitle, midrule, toprule, bottomrule
  },
  fancyvrb=true,
}
\lstdefinestyle{console}{
  style=common,
  backgroundcolor=\color{gray!10},
  aboveskip=5pt,
  belowskip=5pt,
}

\newlist{options}{description}{1}
\setlist[options]{%
  beginpenalty=10000,%
  itemsep=.5\parskip plus .3\parskip minus .2\parskip,
  parsep=.5\parskip plus .3\parskip minus .2\parskip,
  topsep=.5\parskip plus .3\parskip minus .2\parskip,
  partopsep=.5\parskip plus .3\parskip minus .2\parskip,
  style=nextline,labelindent=1em,%
  font=\normalfont\ttfamily}

\colorlet{macro color}{cyan!50!blue!75!black}
\colorlet{option color}{red!50!black}
\colorlet{generic color}{green!40!black}

\newtcolorbox{pseudoTeX}{colback=blue!5,colframe=blue!5,before=\nobreak}
\let\LaTeXorig\LaTeX
\renewcommand\LaTeX{\bgroup\fontfamily{lmr}\selectfont\upshape\LaTeXorig\egroup}

\urldef{\sitewebgind}\url{http://gind.mines-albi.fr}
\urldef{\sitewebminesalbi}\url{http://www.mines-albi.fr}
\urldef{\gmailpaulgaborit}\url{paul.gaborit@gmail.com}
\urldef{\gmailpolimi}\url{name.lastname@polimi.it}
\urldef{\jdmail}\url{j.doe@example.com}
\urldef{\sitex}\url{www.example.com}

\iscramset{
  title={
    TriggerCit: Early Flood Alerting
    using Twitter and Geolocation - a comparison with alternative sources  
  },
  short title={TriggerCit: Early Flood Alerting},
author={
    short name = C. Bono,
    full name=Carlo Bono and Barbara Pernici\thanks{corresponding authors},
    affiliation={
      Politecnico di Milano - DEIB\\
      \href{mailto:name.lastname@polimi.it}{\gmailpolimi}
    },
  },
  author={
    full name= Jose Luis Fernandez-Marquez and Amudha Ravi Shankar,
    affiliation={
      University of Geneva
    },
   },
   author={
    short name = M. O. M\"{u}l\^{a}yim, 
    full name= {Mehmet O\u{g}uz} M\"{u}l\^{a}yim,
    affiliation={
      Artificial Intelligence Research Institute (IIIA-CSIC)
    },
   },
 author={
    full name= Edoardo Nemni,
    affiliation={
      United Nations Satellite Centre (UNOSAT), United Nations Institute for Training and Research (UNITAR)
      },
    },
   }

\usepackage{tikzpagenodes}
\usetikzlibrary{fit}

\begin{document}

\maketitle

\makeatletter
\makeatother

\abstract{

Rapid impact assessment in the immediate aftermath of a natural disaster is essential to provide adequate information to international organisations, local authorities, and first responders.
Social media can support emergency response with evidence-based content posted by citizens and organisations during ongoing events.
In the paper, we propose TriggerCit: an early flood alerting tool with a multilanguage approach focused on timeliness and geolocation. The paper focuses on assessing the reliability of the approach as a triggering system, comparing it with alternative sources for alerts, and evaluating the quality and amount of complementary information gathered. Geolocated visual evidence extracted from Twitter by TriggerCit was analysed in two case studies on floods in Thailand and Nepal in 2021.
}


\section{Introduction}
The increasing use of mobile phones and social media has transformed the way people witness and report climate and weather related disasters. Social media has been demonstrated to be a worthy source of data to provide situational awareness in the course of a disaster event. Effective response to natural hazards, such as floods, can reduce loss of life and mitigate structural damages. Access to timely and accurate data is essential to the disaster response. The effectiveness of the humanitarian response is outside the scope of this work. Instead, we focus on the mechanism of early alert using social media, under the assumption that supporting the activation of the disaster response protocol with pertinent and geolocated evidence is beneficial for the affected community.




Remote sensing data such as satellite images have also proven to be a valuable source of information to support humanitarian relief and disaster response. 
Advances in Artificial Intelligence (AI)
provide a way of extracting relevant information from satellite imagery and thus contribute to speeding up the damage assessment. Even though computer vision techniques can process large amounts of data in a short time, the triggering of such systems is often tied to manual or semi-automated procedures. 

To leverage an alternative and abundant source - social media - this paper proposes a methodology for early flood alerting using Twitter, providing a systematic approach to extract near-real-time and geolocated evidence about onsets of flood events.
The goal of this paper is to analyse how this information correlates with existing alert mechanisms.  



The paper is structured as follows: the next section presents existing disaster alert and trigger mechanisms, and the current use of social media as an early alerting system. In the subsequent section, the TriggerCit methodology for extracting information from social media posts is illustrated. Then, the TriggerCit methodology is applied to two different case studies for flood events in Thailand and Nepal and the results are discussed.

\section{Related work}

\subsection{Existing disaster alert and trigger mechanisms }

\label{sec:existing}Multi-hazard early warning systems and disaster alert mechanisms are a significant component of disaster risk reduction. They help prevent loss of life, reduce the impact of natural hazards, and assist responders in disaster handling. In this section, we will discuss two key state-of-the-art alert mechanisms.  

\emph{The Global Disaster Alert and Coordination System}\footnote{\url{https://www.gdacs.org}} (GDACS), started in 2004, is a joint initiative of the United Nations Office for the Coordination of Humanitarian Affairs (UN OCHA)\footnote{\url{https://www.unocha.org}} and the European Commission (EC). It is designed to alert the international community during sudden-onset disasters that might require international assistance. GDACS serves to consolidate and improve the dissemination of disaster-related information and coordination in the first phase of a disaster. The GDACS website consists of different tools, which have been incorporated into disaster response plans by several governments and organisations. The GDACS Disaster Alerts are auto-generated based on algorithms and available data to alert oncoming disasters. The system calculates alerts based on specific risk indicators, like the severity of the disaster, the affected population, and the country's vulnerability. Subscribers are notified via Email and SMS (\cite{kavallieros2015ppdr}). GDACS information is openly accessible. For the first phase of a disaster, governments and disaster response organisations rely on GDACS alerts and automatic impact estimations to plan international assistance. The Virtual On-Site Operations Coordination Center (Virtual OSOCC)\footnote{\url{https://vosocc.unocha.org}} facilitates information exchange and coordination.

\emph{The Global Flood Awareness System}\footnote{\url{https://www.globalfloods.eu}} (GloFAS) is a component of the Copernicus Emergency Management Service (CEMS) that delivers global hydrological forecasts to registered users. GloFAS combines information from the satellites and forecast models to produce GloFAS forecasts, Seasonal forecasts, and Impact Forecasts (\cite{kavallieros2015ppdr,zajac2014recent}).
The GloFAS Forecasts are available free of cost to all, with registered users from Hydro-Meteorological services, academia, and the humanitarian sector (\cite{HydroCon54:online}).

\subsection{Social media as early warning and alerting systems}
Several social media sources have been considered in the literature, the main one being Twitter\footnote{\url{https://twitter.com}} as it provides public application programming interfaces (APIs) to search posts and related metadata. Several research works have been published on tweets analysis. Here we focus on tweet analysis for early event detection. One of the first papers on the topic is \cite{DBLP:journals/tkde/SakakiOM13}, in which the focus is on the timeliness of the extracted information, in particular in the case of earthquakes. Like in other papers analysing social media, the post geolocation is extracted from Twitter metadata. As studied in the literature (e.g., \cite{scalia2021cime}), this is usually available only in a small percentage of the posts.
Recent papers focused on analysing in parallel several sources of information for deriving awareness on emergency events. In \cite{shoyama21}, the impact of extensive floods in Japan has been assessed using hydrogeological information and tweets for detecting event outbreaks and phases in the affected areas.

The importance of images and geolocation via text has been recently emphasised to overcome the limitations of native Twitter geolocations. There are several advantages of having precisely geolocated tweets.
\cite{fohringer2015social} show how a very limited number of tweets, combined with elevation profiles, can give a precise map of flooded areas; also \cite{scotti2020enhanced} emphasise the integration of tweet data and hydraulic modeling for mapping flooded areas. \cite{9091866} uses text to geolocate posts and to classify phases of events automatically. The geolocation steps are based on Named-Entity Recognition (NER) and gazetteers to identify locations. However, the focus is mainly on locations in the US and the English language. 

A more comprehensive approach to social media posts geolocation has been proposed in the CIME system (\cite{scalia2021cime}). The NER is based on Polyglot (\cite{al2015polyglot}), which supports more than 40 languages. CIME disambiguates candidate locations using Nominatim\footnote{\url{https://nominatim.openstreetmap.org}}, a search engine for OpenStreetMap data. Disambiguation considers candidate locations, their distances, and their ranks in the OpenStreetMap administration level hierarchy.

Post geolocation can achieve a significant, although variable, level of precision. CIME was exploited in the E2mC  project (\cite{havas2017e2mc}) to support rapid mapping activities within Copernicus Emergency Management Services (EMS)\footnote{\url{https://emergency.copernicus.eu}}. An important issue that emerged is the high number of duplicate and non-relevant images.
A systematic approach to image filtering for geolocated tweets has been proposed in 
\cite{DBLP:conf/icse/NegriSARSSFCP21}, developing the VisualCit image processing toolkit, aimed at extracting indicators about emergency events. The toolkit allows image filtering based on contents, such as similarity, non-photographic content, presence of specific objects, and event-specific classifiers (including floods, see \cite{genderandfloods21}).
Other authors have proposed to create event-specific classifiers (e.g., \cite{IMRAN2020102261}) automatically, without focusing on locations and systematic data preprocessing.

Another important issue is determining the severity of events using tweets as a source of citizen knowledge. A combination of sentiment analysis and frequency of impact-related keywords has been used in \cite{kankanamge2020determining}.

The availability of a representative set of search keywords is paramount for leveraging social media data. The goal is to get sufficient coverage of the event, both in the initial phase and as the event unfolds, as shown in \cite{Olteanu_Castillo_Diaz_Vieweg_2014}. 
In several papers, general purpose keywords are used. A systematic approach has been developed in E2mC based on data mining techniques for multilingual keyword extraction (\cite{kirsch2018robust}).

In the present paper we propose TriggerCit, a systematic methodology to extract flood alerts from Twitter posts. We focus on detecting the occurrence of flood events within the first 48 hours of the event, and describing them at the regional administrative level, providing image evidence with detailed locations. The goal is to select localised, relevant information from social media in a short time.
The approach leverages VisualCit processing and CIME geolocation. The results are compared, over two case studies, with data from the event alert systems described in \ref{sec:existing}.




\section{Methodology} \label{sec:method}

\begin{figure}[h!]
\centering
  \includegraphics[width=0.9\columnwidth]{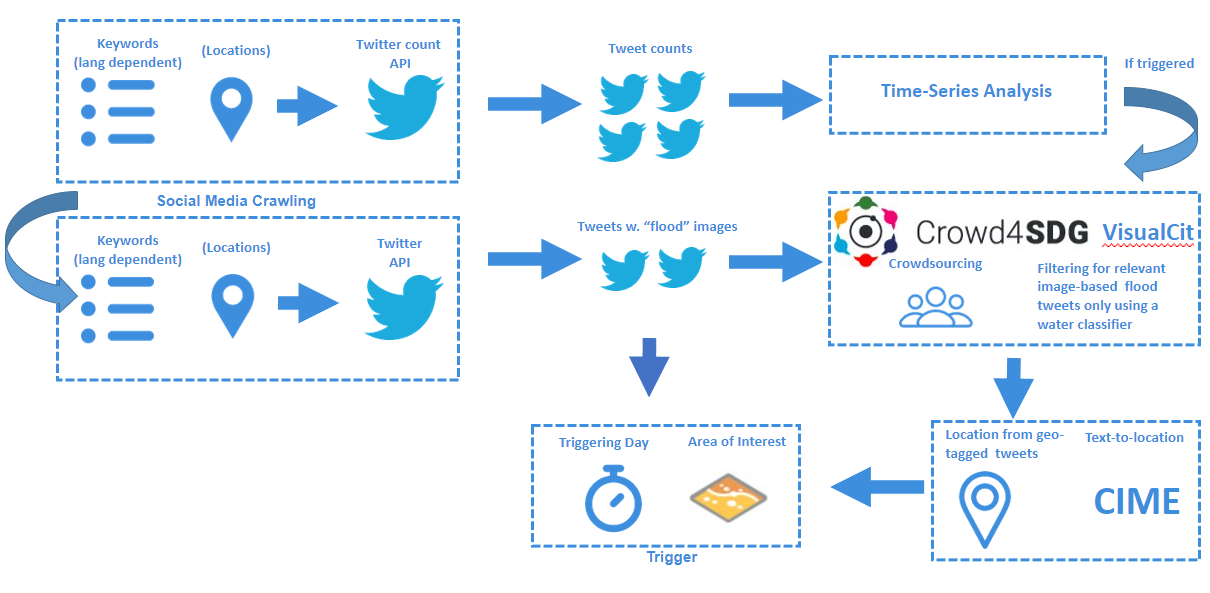}
  \caption{TriggerCit methodology for identifying  occurrences of flood events and related locations.}
    \label{fig:pipeline}
\end{figure}

The proposed approach, shown in Fig.~\ref{fig:pipeline}, is implemented as an analytics pipeline aimed at detecting flood events and extracting high-quality event-related information. The components are adapted to Twitter. However, most of the concepts can be generalised to other compatible social media.

The main steps of TriggerCit are illustrated in the following.
\begin{enumerate}
  \item \emph{Word dictionaries for social sensing}\newline 
 A \emph{seed dictionary}, which consists of a curated dictionary of words, is built for each language and event of interest. Ideally, this dictionary should be rich enough to account event variability, yet specific enough to match relevant messages only. Language-dependent dictionaries should be curated by expert native speakers. Moreover, the same event could be described in different languages, since multiple languages could be spoken in affected regions. In the current experiments, we assume that a small general-purpose set of terms is available, as we provide novel techniques for augmenting the event-related set of words (see Step~7).
  
  \item \emph{Word count monitoring for event detection} \newline
  With proper linguistic resources in place, a monitoring system for disaster-related terms can be implemented. Dictionaries can be used to build one or more queries towards this goal. We chose to build a query as the logical OR of the words in the dictionary. Alternatively, per-word time series can be used. Getting the time series of the number of matched posts is straightforward with Twitter APIs. The main advantage of querying counts is that no actual tweet content data is exchanged, so almost no computation is involved at this stage. Moreover, no tweet quota is consumed, which could be critical if the number of queries to monitor is high (e.g., many countries/languages). Rate limits are compatible with near-real-time monitoring. Geographical boundaries can be specified: however, this will limit the results to the few tweets that are natively geotagged in Twitter, which is usually a limiting factor as discussed in \cite{scalia2021cime} and other works.
  
  \item \emph{Triggering} \newline
  Using the counts time series, an event triggering system can be designed. Ideally, the component should be a supervised classifier, using one or more validated datasets such as historical GDACS alerts for the event starts (and ends) as reference data. A more straightforward mechanism triggers an event if the tweet count growth exceeds some threshold.
  
  \item \label{step:query} \emph{Get tweet contents} \newline
  If an event is likely occurring, we selectively download tweet data for the time interval of the event using the Twitter search endpoint, excluding retweets. Full-archive search API is available for research purposes. The recent search API is also available, which is accessible to the general public and limited to the last 7 days.
  
  \item \emph{Extract relevant content with VisualCit} \newline
   We then concentrate on tweets with images, applying a VisualCit semantic filtering pipeline to extract relevant contents.
   The tasks in the pipeline are ordered ``quickest first'' as a heuristic to get a faster overall processing time.
   The proposed pipeline consists of the following tasks:
  
    \begin{enumerate}
         \item Remove duplicated images and similar images (e.g., rescaled versions, modified versions).
         \item Remove non-photos, such as drawings, screenshots and computer-generated images.
         \item Remove not-safe-for-work images.
    \end{enumerate}
    
    The confidence thresholds for the single tasks are chosen through empirical validation, to achieve a reasonable performance without substantially removing useful data.

  \item \emph{Use CIME for geolocation} \newline
  The CIME geolocation algorithm described in \cite{scalia2021cime} is then applied to associate tweets with geographical entities, whenever possible. Locations in unrelated countries are filtered out using national administrative boundaries.
  
  \item \emph{Use pipeline output as feedback for input}\label{step:promising} \newline
     Having a selected set of relevant items could be useful to leverage the significant, unfiltered mass of tweets without images. A straightforward approach could be extracting tweets related to the ones in the VisualCit output, searching in the broader set of text-only tweets retrieved in Step~\ref{step:query}.
    Textual frequency analysis may help us to identify keywords that were not explicitly listed in the original seed dictionary. New keywords can then be posed as a query to the corpus of text-only tweets. A simple measure like the TF-IDF score (possibly combined with the widespread cosine similarity) can rank individual tweets concerning this query. Selected text-only tweets with a score higher than some threshold (henceforth referred to as \emph{promising text-only tweets}) can then be fed to CIME for geolocation, thus expanding the real-time geographical description of the event. In next section,
    focusing on the use case in Thailand, we show how this approach can effectively augment the map of relevant tweets. This approach could be further extended in order to refine the input space by recovering information with additional queries in Step~\ref{step:query}.
  
   \item \emph{Visualization} \newline
   The process output consists of a representative set of social media posts, used to deliver thematic maps providing geolocated evidence to stakeholders.

\end{enumerate}

\section{Case studies}

In this section, we apply the proposed methodology to two flood events that occurred in 2021 in Thailand and Nepal.
The two events were selected since the United Nations Satellite Centre (UNOSAT)\footnote{\url{https://unitar.org/sustainable-development-goals/united-nations-satellite-centre-UNOSAT}} supported both activations in Nepal\footnote{\url{https://unitar.org/maps/countries/70}} and Thailand\footnote{\url{https://unitar.org/maps/countries/100}} with satellite-derived maps and AI-based flood detection analysis, updating operational dashboards in near-real-time\footnote{\url{ https://unitar.org/about/news-stories/news/unosat-flood-ai-dashboards-nepal-creation-one-stop-shop-real-time-evidence-based-decision-making}} and providing daily updates.  

\subsection{Thailand Flood Event - September-October 2021} \label{sec:case-thai}

\subsubsection{Event description}
In late September and October 2021, tropical storm Dianmu inundated regions in Vietnam, Laos, Cambodia, and Thailand. According to the UNOSAT Thailand flood monitoring dashboard\footnote{\url{https://unosat-geodrr.cern.ch/portal/apps/opsdashboard/index.html\#/4f878691713a40f3b8ef3140e63c9f6d}}, around 14.3 square kilometers of the country were inundated between September 27 and October 27, 2021. An estimated 1.4 million people were affected by the flooding, with 32 of 76 provinces affected by heavy rains for nearly a month. The statistics mentioned above were computed in the aftermath of the flood events and should be considered as a preliminary analysis since they have not been validated yet in the field.

\subsubsection{Data description}
\begin{figure}[th!]
    \centering
    \subfigure[Thai entries]{\includegraphics[width=0.75\textwidth]{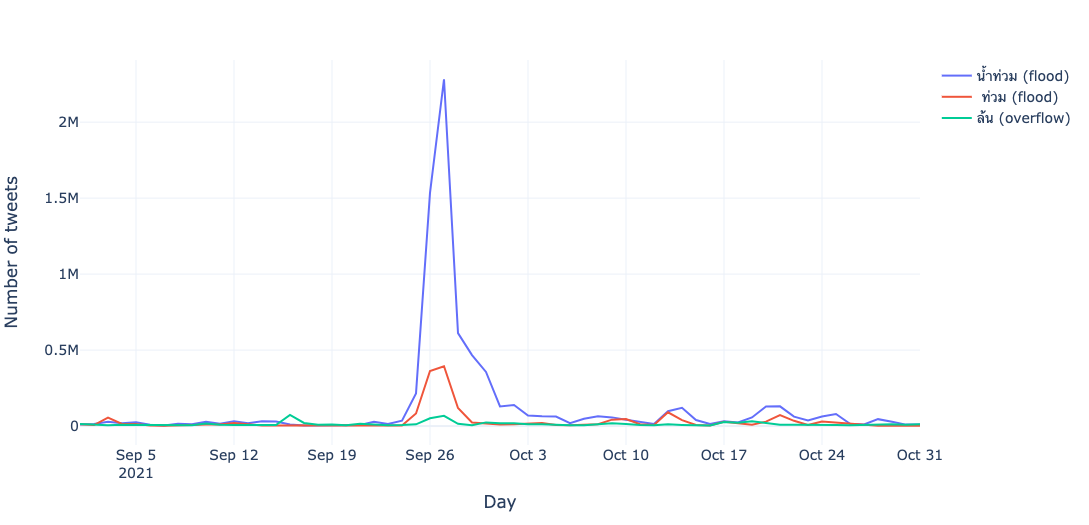}}
    \subfigure[Thai / English]{\includegraphics[clip,trim=0 0 0 1.2cm,width=0.75\textwidth]{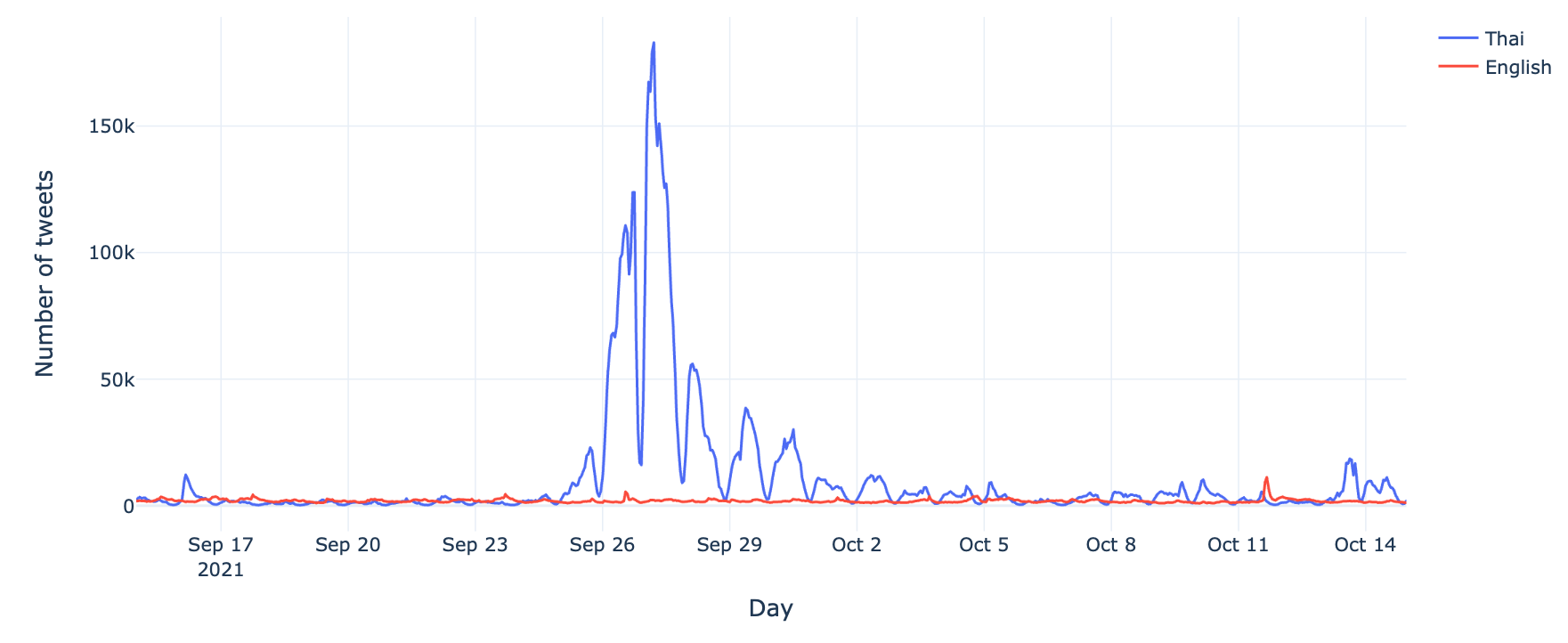}}
    \caption{Daily tweet counts for Thai dictionary entries}
    \label{fig:thai_counts_2}
\end{figure}

We queried Twitter with the logical OR of the three most frequent terms in the dictionary (see Fig. \ref{fig:thai_counts_2}(a)) of Thai terms correlated to flood events and provided by the UNOSAT office in Bangkok. The event onset is clear using the Thai dictionary, but not with the equivalent English dictionary (see Fig. \ref{fig:thai_counts_2}(b)).

We concentrated the analysis on the 48 hours between September 26, 00:00 UTC and September 28, 00:00 UTC, since there is marked evidence of the event onset. Over this period, more than 4 million tweets match the seed query.


\begin{table}[ht]
    \centering
    \begin{tabular}{||c c||} 
     \hline
     \bf{Data} & \bf{Count}\\ [0.5ex] 
     \hline\hline
     All tweets, September 26-27 & 4'145'447 \\ 
     \hline
     No retweets & 66'868 \\
     \hline
     Containing images & 6'292 \\
     \hline
     Native Twitter locations & 227 \\
     \hline
     Overall images & 8'774 \\
     \hline
     Passed VisualCit filters & 3'056 \\
     \hline
     Places geolocated by CIME & 1'671 \\
     \hline
 
    \end{tabular}
    \caption{Dataset cardinality through processing, Thailand case study}
    \label{tab:thai_count}
\end{table}

\begin{table}[ht]
    \centering
    \begin{tabular}{||c c c||} 
     \hline \hline
     \bf{Level} & \bf{Admin level} & \bf{Count}\\ [0.5ex] 
     \hline\hline
     Province & 4 & 7 \\ 
     \hline
     District & 6 & 8 \\ 
     \hline
     Municipality or subdistrict & 8 & 176 \\ 
     \hline
     Village or community & 10 & 9 \\ 
     \hline
     Other points  & 15 & 1'265 \\ 
     \hline
 
    \end{tabular}
    \caption{Administrative level counts for CIME geolocations}
    \label{tab:cime_adm_count}
\end{table}

\begin{figure}[th!]
    \centering
    \subfigure[Removed items]{\includegraphics[width=0.45\textwidth]{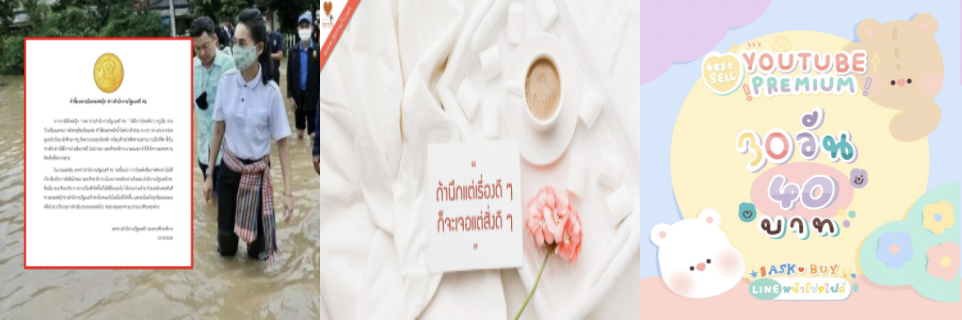}}   
    \subfigure[Kept items]{\includegraphics[width=0.45\textwidth]{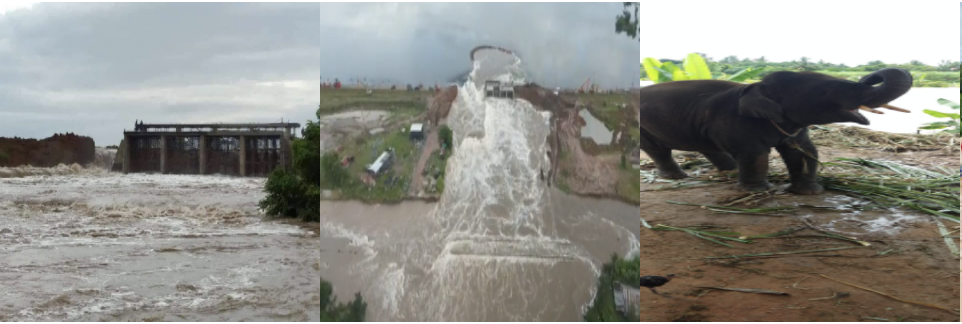}} 
    \caption{Example images for (a) items being removed by the pipeline, and (b) items being kept}
    \label{fig:thai_filtered}
\end{figure}

\textit{Processing output}\newline
 We first focused on processing the 8'774 images available\footnote{The datasets for this paper will be made available on Zenodo.} with VisualCit (see Fig. \ref{fig:thai_filtered} for an example of the result). The VisualCit pipeline is configured to remove duplicates, non-photos, and not-safe-for-work contents. The average processing time for each item, including network overheads, is 171ms (roughly 20'000 elements/hour) on a single server.

\begin{figure}[th!]
    \centering
    \subfigure[Native Twitter geolocation]{\includegraphics[width=0.245\textwidth]{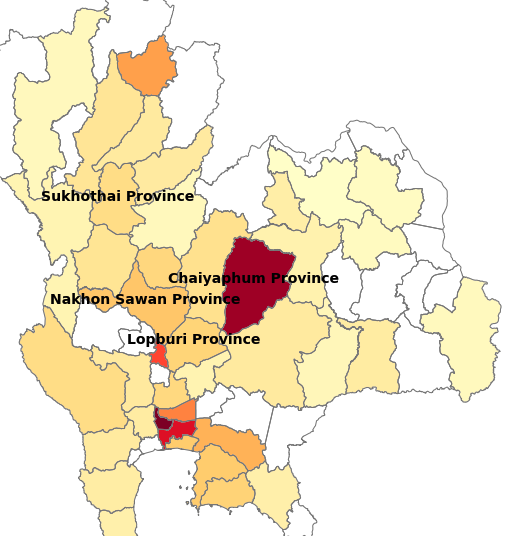}}   
    \subfigure[VisualCit geolocation]{\includegraphics[width=0.245\textwidth]{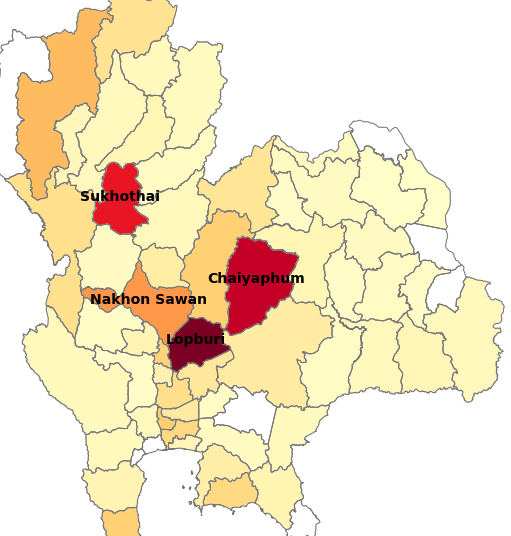}} 
    \subfigure[VisualCit geolocation with extended tweet set]{\includegraphics[width=0.245\textwidth]{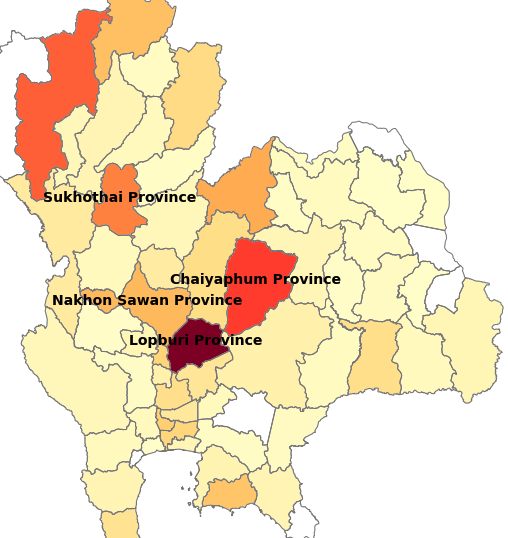}}
    \subfigure[Affected people]{\includegraphics[width=0.245\textwidth]{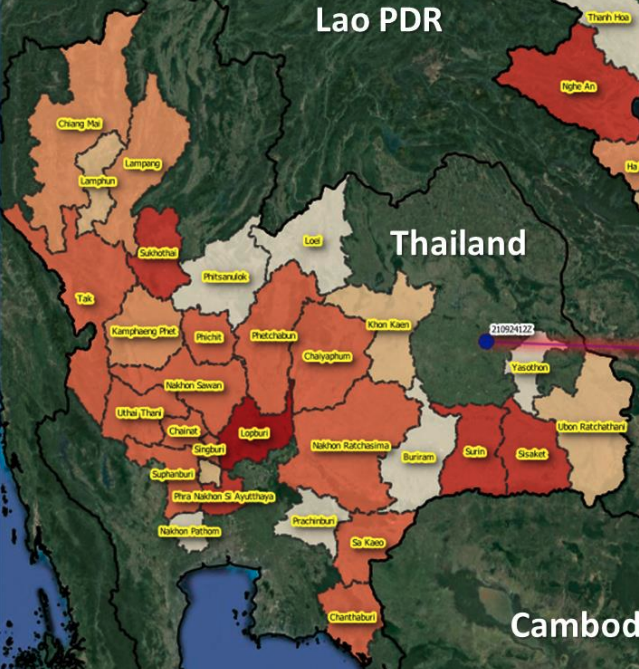}}
    \caption{Geolocations / inhabitants ratio by region (a) Twitter native geolocations, (b) Twitter native + CIME geolocated, (c) Twitter native + CIME geolocated from extended dataset (with images + promising text-only tweets), and (d) Number of affected persons by region at September, 28th (source: ReliefWeb)}
    \label{fig:thaigeo}
\end{figure}

The resulting dataset is then processed with the CIME algorithm in order to extract locations. The distribution of the places extracted by CIME does not simply replicate the distribution of the native geotags, as it can be seen comparing Fig. \ref{fig:thaigeo}(a) and Fig. \ref{fig:thaigeo}(b), and allows a clearer description of affected areas.
In our experimental setup, to the 206 locations natively present in the tweets and located in Thailand, we can add 1'465 locations that are within Thailand state borders.
The average processing time for each item with CIME is 224ms\footnote{With no parallelism}. The correlation between the geolocated data and the affected areas, as described by official sources\footnote{\url{https://reliefweb.int/sites/reliefweb.int/files/resources/FlashUpdate_02_28Sep2021-TC-DIANMU-THLVNMLAO.pdf}}, can be appreciated in Fig. \ref{fig:thaigeo}. Population estimates by province used for data normalisation were extracted from Wikipedia\footnote{\url{https://en.wikipedia.org/wiki/Provinces_of_Thailand}}.
Mainly, specific locations or entities with sub-district administrative level are extracted (see Table \ref{tab:cime_adm_count}\footnote{OpenStreetMap administrative levels for Thailand \url{https://wiki.openstreetmap.org/wiki/Tag:boundary=administrative}.}). Data cardinalities through processing steps are described in Tab. \ref{tab:thai_count}.

Fig.~\ref{fig:thaigeo}(c) shows how we can further extend location analysis by exploiting the text-only tweets as described in Step~\ref{step:promising} of the TriggerCit methodology.
A word frequency analysis of the VisualCit output provided us with two new keywords (water and house). We posed these two together with the original seeds as a query to the text-only corpus and measured the cumulative TF-IDF score of each tweet. We regarded the 6'021 tweets within the 0.9 percentile of the score as \emph{promising} and fed them to CIME for geolocation. 661 of these tweets were successfully geolocated within Thailand by CIME. 

\subsubsection{Assessment of the results}

TriggerCit outputs an alert on September 27 due to the increase of tweets compared to the days before considered as a baseline. It also returns a weighted set of administrative levels where the alert is coming from, which partially correlates with the number of affected people and gives a preliminary indication of the affected regions.

\subsection{Nepal Flood Event - June-July 2021}
\subsubsection{Event description}

Between June and October 2021, 673 people have lost their lives in Nepal due to floods and landslides caused by seasonal and unseasonal heavy rain and severe weather\footnote{\url{https://reliefweb.int/disaster/fl-2021-000134-npl\#overview}}. Among the events at the beginning of the monsoon season, heavy rain triggered multiple landslides and flooding on the Melamchi river, in the Sindhupalchok district, resulting in more than 20 deaths\footnote{\url{https://en.wikipedia.org/wiki/2021_Melamchi_flood}}. Damage and casualties followed since late June and the beginning of July, marking the onset of the disaster event. The case study focuses on evidence from tweets gathered in this early stage.

\subsubsection{Data description}
\begin{figure}[th!]
    \centering
    \subfigure[Nepali entries in June]{\includegraphics[width=0.75\textwidth]{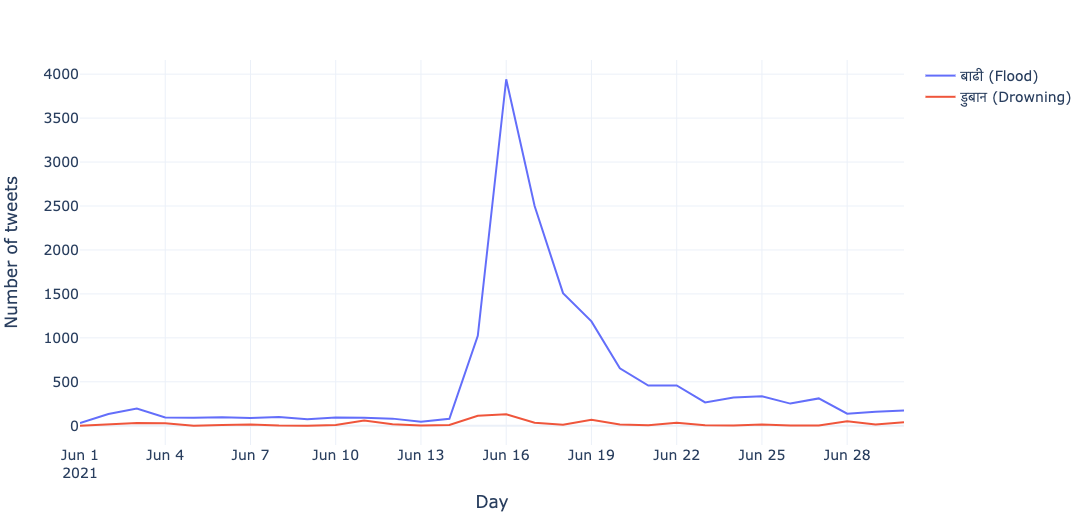}}
    \subfigure[Nepali entries in July]{\includegraphics[width=0.75\textwidth]{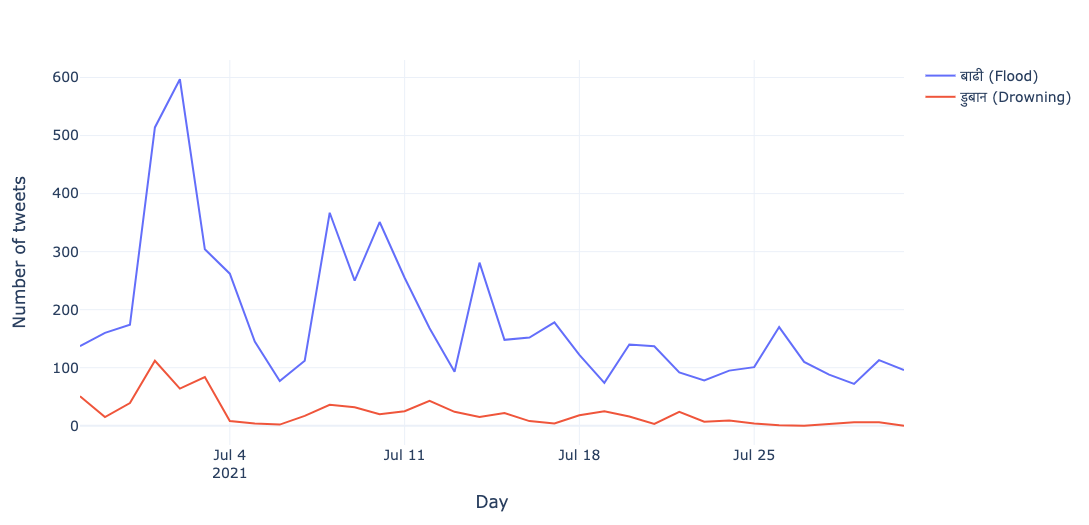}}
    \subfigure[Nepali / English]{\includegraphics[clip,trim=0 0 0 1.2cm,width=0.75\textwidth]{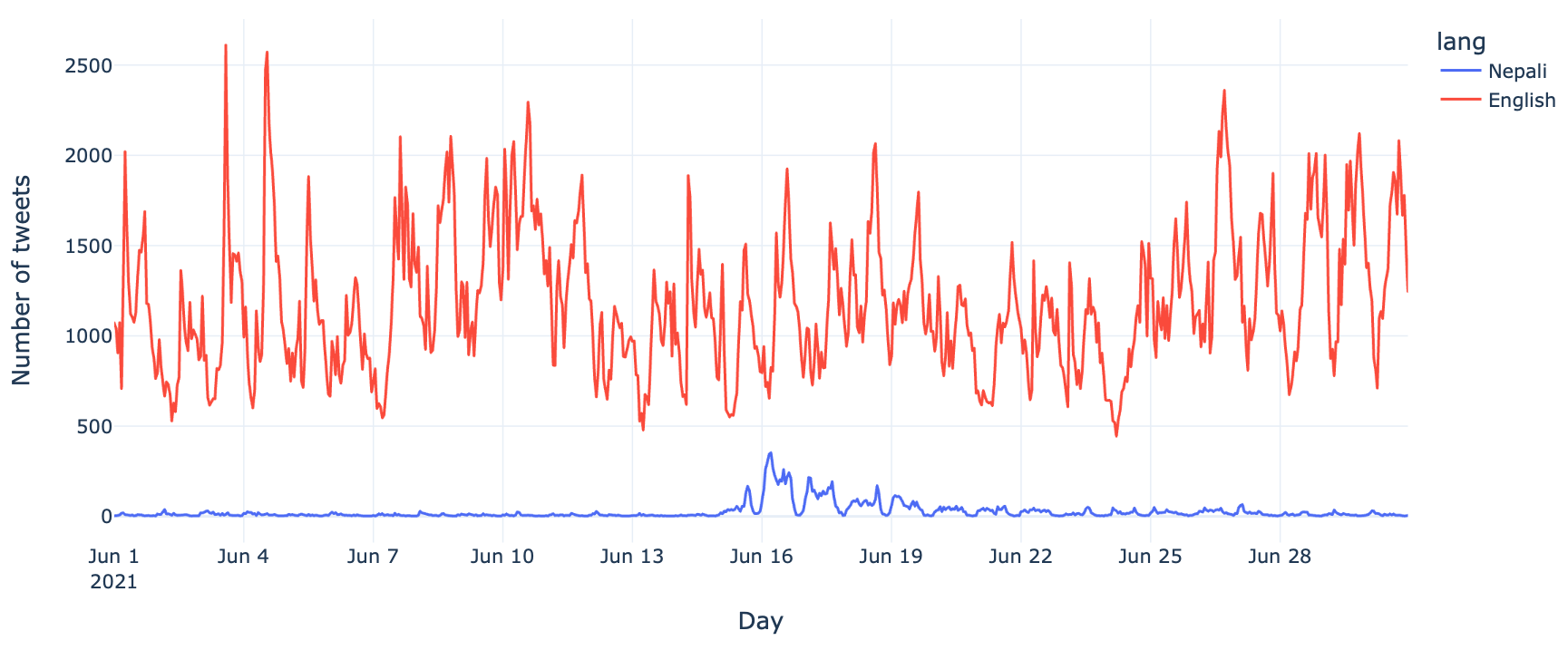}}
    \caption{Daily tweet counts for Nepali dictionary entries}
    \label{fig:nepali_counts_2}
\end{figure}

Using a pre-defined dictionary of Nepali words generically related to flood events, the query consisted of the logical OR of the terms (see Fig. \ref{fig:nepali_counts_2}). As in the case of Thailand floods, it is difficult to extract evidence with English seed keywords (see Fig. \ref{fig:nepali_counts_2}(c)). 


The most evident event using our dictionary happened in the middle of June, as seen in Fig. \ref{fig:nepali_counts_2}(a), which coincides with the event dates reported in the news. Over June 16-17, 6’639 tweets matched the query. It is relevant that, among the few posts with native Twitter geolocations, none contained information specific to this event.

We replicated the analysis for the first two days of July, since in these days several incidents with casualties happened in different places of Nepal\footnote{\url{https://floodlist.com/asia/nepal-floods-landslides-july-2021}}. In this case, the time series of tweets matching our query show an increasing trend, albeit less evident than in the previous case.

\begin{table}[ht]
    \centering
    \begin{tabular}{||c || c || c ||} 
     \hline \hline
     \bf{Data} & \bf{ June 16-17} & \bf{July 1-2}\\ [0.5ex] 
     \hline\hline
     All tweets & 6'639 & 1'225\\ 
     \hline
     No retweets & 2'807  & 594 \\
     \hline
     Containing images & 261 & 63 \\
     \hline
     Native Twitter locations & 8 & 10 \\
     \hline
     Overall images & 391 & 80 \\
     \hline
     Passed through VisualCit & 218 & 55 \\
     \hline
     Places geolocated with CIME & 51 & 10 \\
     \hline
 
    \end{tabular}
    \caption{Data cardinality in Nepal case study for June 16-17 and July 1-2}
    \label{tab:nepal_count}
\end{table}

\subsubsection{Processing output}
The experimental setup is equivalent to the one used for the Thailand case study. Data cardinalities through processing are reported in Tab. \ref{tab:nepal_count}. Since the pipeline did not support Nepali natively, we processed Nepali texts with a publicly available NER tool for Nepali\footnote{\url{https://github.com/oya163/nepali-ner}}, described in \cite{8998477}. The processed posts are then fed to a modified CIME algorithm for location extraction. 

\begin{figure}[h!]
    \centering
    \subfigure[Nepali tweets geolocated with CIME - June 16-17, 2021 ]{\includegraphics[width=0.45\textwidth]{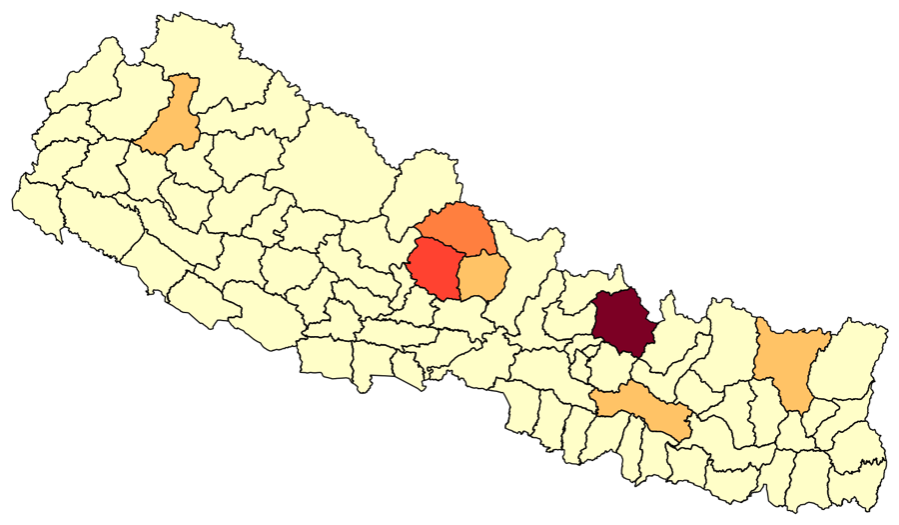}}
        \subfigure[GloFAS reporting points and Rapid Impact Assessment map - June 17, 2021]{\includegraphics[width=0.45\textwidth]{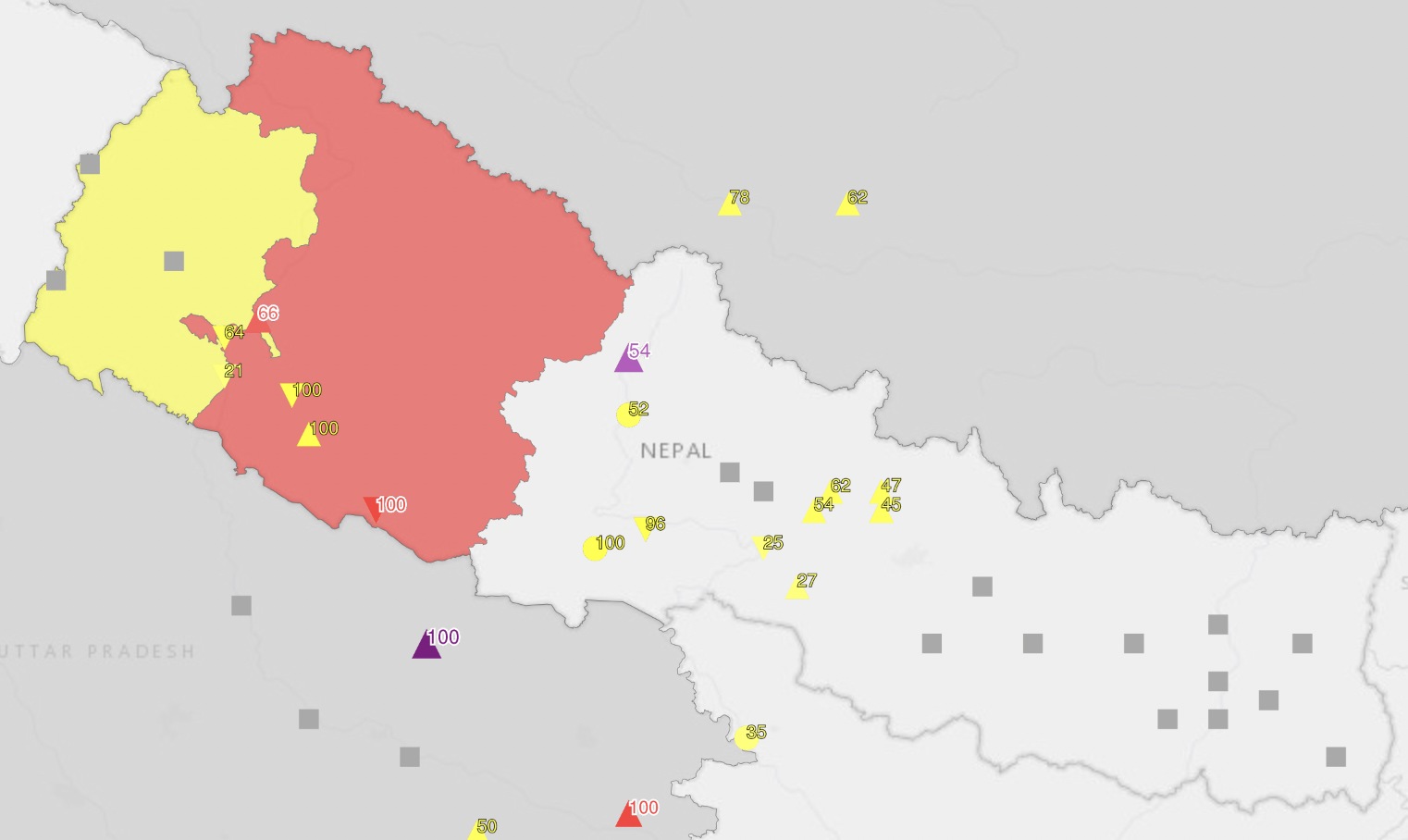}}
    \caption{Nepali case study - June 16-17, 2021}
    \label{fig:nepali_counts_map}
\end{figure}

For data in June 16-17, among the 51 locations extracted, 21 refer to the Bagmati province, in which the event is taking place, and 19 explicitly refer to Sindhupalchok district (as Fig. \ref{fig:nepali_counts_map} shows). Another 11 generically refer to Nepal, so they are ignored. It is interesting to note that, compared to the GloFAS Rapid Impact Assessment map, there is a complementarity regarding this event, since there are no reporting points for the Sindhupalchok district. 

\begin{figure}[h!]
    \centering
    \subfigure[UNOSAT Flood AI Monitoring Dashboard - July 1-2, 2021]{\includegraphics[width=0.45\textwidth]{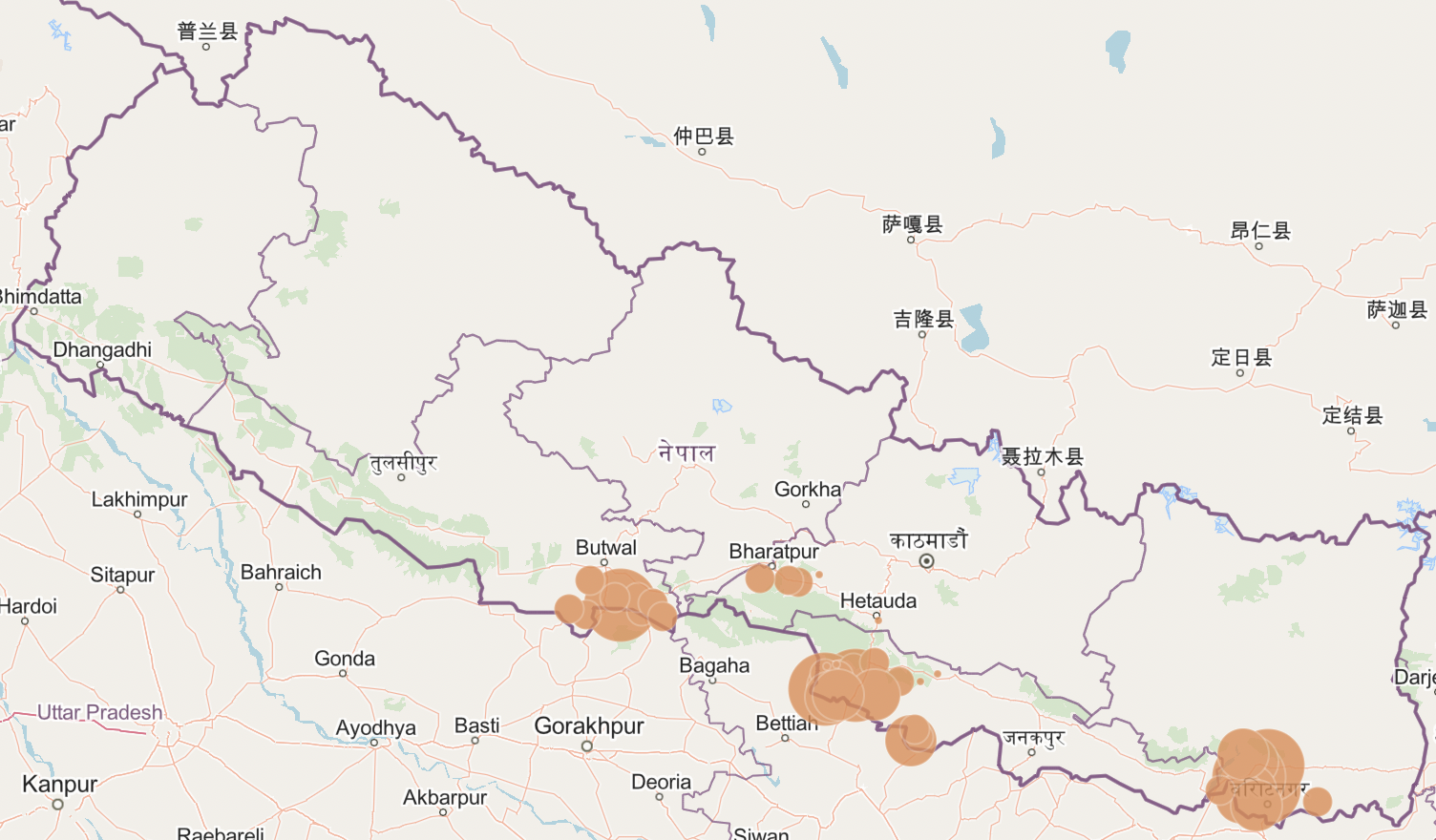}}
    \subfigure[GloFAS reporting points and Rapid Impact Assessment map - July 2, 2021]{\includegraphics[width=0.45\textwidth]{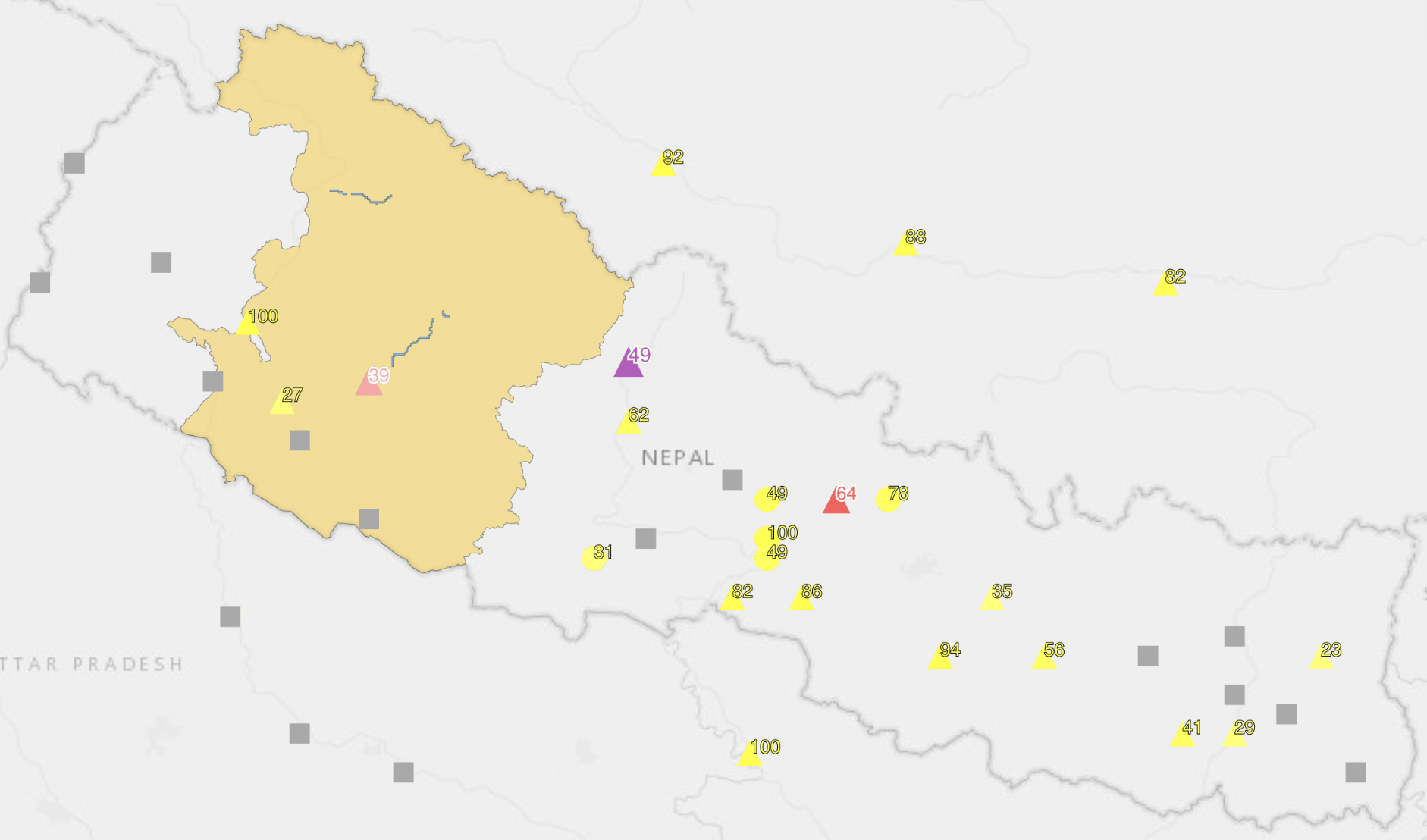}}
    \subfigure[Geolocated Nepali tweets by district - July 1-2, 2021 ]{\includegraphics[width=0.45\textwidth]{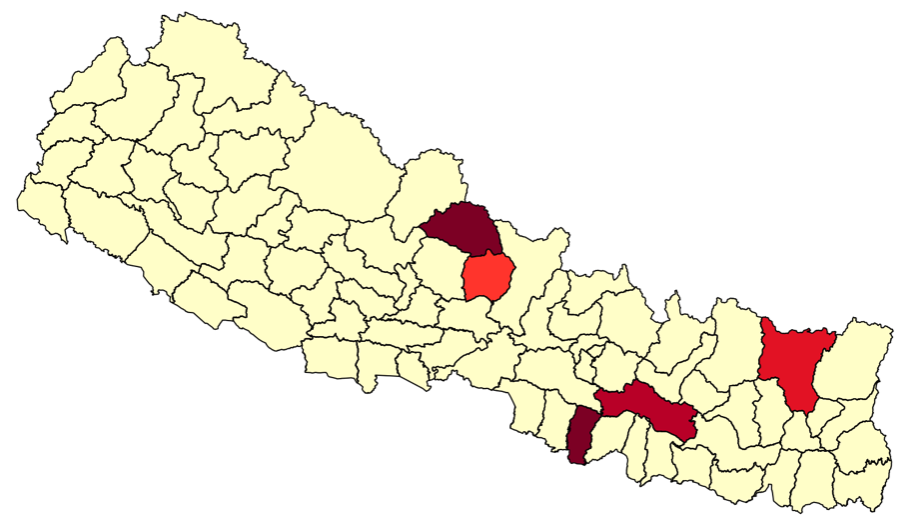}}
    \subfigure[Nepal Disaster Risk Reduction portal incident count by district - July 1-2, 2021]{\includegraphics[width=0.45\textwidth]{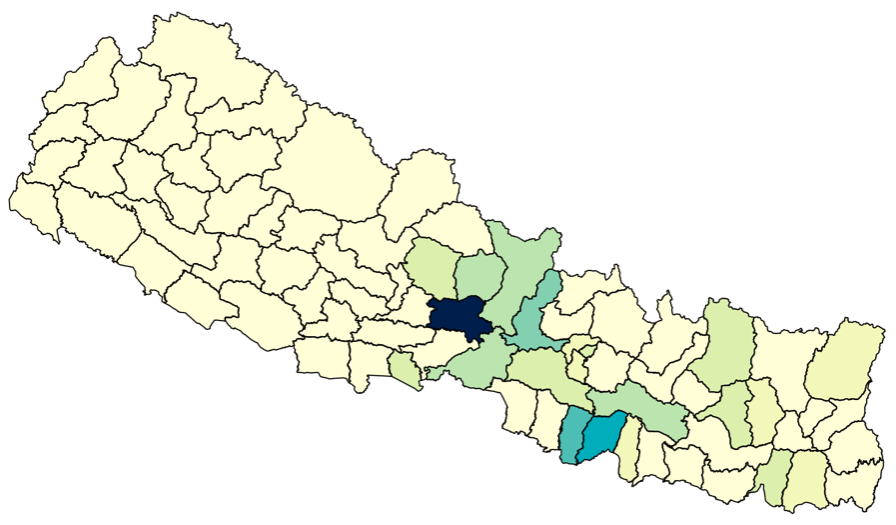}}
    \caption{Nepali case study - July 1-2, 2021}
    \label{fig:nepali_counts_map_july}
\end{figure}

For data in July 1-2, an overview of the processed data is reported
in Table \ref{tab:nepal_count}. The amount of evidence gathered is smaller than the previous time frame, with only 6 useful locations extracted after running the images through the pipeline (4 out of 10 are generic mentions of Nepal), and 14 in the overall set of tweets with images. Nevertheless, it is worth comparing the results with different sources of information to understand how to complement available information, which is not always concordant and could suffer from data scarcity. In Fig. \ref{fig:nepali_counts_map_july} we gathered figures for different signals on July 2, coming from UNOSAT, GloFAS, TriggerCit, and reports from the Nepal Disaster Risk Reduction portal (NDRR)\footnote{Events of flood, heavy rain and landslides reported from \url{http://drrportal.gov.np/} for July 1-2}. The relevance of the locations extracted with our approach is partial, but significant in some cases. The limited amount of data is possibly related to the specificity of our queries, which are focused on flood terms, while 50 out of 59 incidents reported by NDRR have a ``Heavy Rainfall'' or ``Landslides'' label. The impact of a tailored dictionary should be further investigated.

\subsubsection{Assessment of the results}
TriggerCit outputs an alert on June 16 due to the increase of tweets compared to the days before considered as baseline (Fig. \ref{fig:nepali_counts_map}). This trigger is related to a specific local event. Heavy rainfall caused a number of floods and landslides, across different regions of Nepal during the following weeks. Comparing the two periods, it appears that the peaks of tweets are more likely for sudden unexpected events, rather than for an expected event which is developing over time with many local events over an extended area.
The corresponding maps and reference maps are shown in Fig. \ref{fig:nepali_counts_map_july}. The ability of TriggerCit of keeping track of these events is still under evaluation.

\section{Comparison with alternative sources}

UNOSAT Rapid Mapping Service\footnote{\url{https://www.unitar.org/maps/unosat-rapid-mapping-service}} provides satellite image analysis during humanitarian emergencies, both natural disasters, and conflict situations. With a 24/7 year-round availability to process requests, a team of experienced analysts ensures timely delivery of satellite imagery-derived maps, reports, and data ready for direct inclusion in Geographic Information Systems (GIS) according to needs. 
The service can be activated by UN offices and agencies Government agencies, the Red Cross and Red Crescent Movement (ICRC and IFRC), international and regional organizations, and Humanitarian Non-Governmental Organizations (NGOs). As shown in Table \ref{tab:comparison}, UNOSAT was activated on September 28, 2021 for the flood event in Thailand and on June 30, 2021 for the flood in Nepal. The activation data refers to the day when UNOSAT received the request to support the flood events. 
We were not able to find any flood activation from Copernicus EMS\footnote{\url{https://emergency.copernicus.eu/mapping/\#zoom=2\&lat=27.6533&lon=-25.0083\&layers=0BT00}} in Thailand and Nepal respectively. 

GDACS reported floods on September 27, 2021 for Thailand\footnote{\url{https://www.gdacs.org/report.aspx?eventtype=FL&eventid=1101113}}, and on June 28, 2021 for Nepal\footnote{\url{https://www.gdacs.org/report.aspx?eventid=1100949&episodeid=1&eventtype=FL}} both with a green GDACS score\footnote{\url{https://www.gdacs.org/Knowledge/models_FL.aspx}} of 0.5. The event on June 16, 2021 was not recorded in GDACS following floods reported on the GDACS website\footnote{\url{https://www.gdacs.org/flooddetection/currentfloods.aspx}}.
We found the events reported by GloFAS using the GloFAS Map Viewer\footnote{\url{https://www.globalfloods.eu/glofas-forecasting/}} on September, 24 in Thailand with a yellow and red score of Rapid Impact Assessment and on June 28, 2021 in Nepal with a orange score of Rapid Impact Assessment. FloodList also reported news of flood in Thailand\footnote{\url{https://floodlist.com/asia/thailand-tropical-storm-dianmu-floods-september-2021}} on September 27, 2021, and in Nepal on July 4, 2021\footnote{\url{https://floodlist.com/asia/nepal-floods-landslides-july-2021}} and on June 16, 2021\footnote{\url{https://floodlist.com/asia/bhutan-and-nepal-flash-floods-leave-at-least-10-dead-many-missing}}. 

In the two scenarios, TriggerCit respectively returned a large scale and a local scale alert, both in a timely manner and accompanied by a valid geographical description, while providing information complementary to existing disaster alert mechanisms (e.g., images).

\begin{table}[h!]
\centering
\begin{tabular}{|c|c|c|}
\hline \hline
\bf{Dates / Countries }           & \bf{Thailand }                & \bf{Nepal}                    \\ \hline
UNOSAT activation            & 28/09/2021               & 30/06/2021               \\ \hline
Copernicus EMS activation              & None                     & None                     \\ \hline
GDACS Disaster Alerts &     27/09/2021 (Green Alert)  & 28/06/2021 (Green Alert) \\ \hline
GloFAS                       & 24/09/2021              & 28/06/2021     \\ \hline
FloodList reported news                 & 27/09/2021               & 04/07/2021               \\ \hline
TriggerCit                   & 26/09/2021               & 02/07/2021                     \\ \hline
\end{tabular}
\caption{Timeline of alerts for the Thailand and Nepal activations}
\label{tab:comparison}
\end{table}

\section{Limitations and Future Work}

Linguistic limitations while mining multi-lingual tweets are mentioned in most of the works discussed in the state-of-the-art. The task is usually easier if the tweets are in English, given the quality of linguistic resources and the number of posts available. The use of native languages for
communication has always posed a significant challenge to social media mining. In this paper, we propose a multi-language and extensible approach towards a solution to this problem. Further work is needed to develop fast and reliable seed dictionaries for social media monitoring with TriggerCit. For this purpose, several directions have to be investigated, leveraging on local resources, including local administrations and NGOs, and citizens. They can provide information about the nature of repeating local events, such as monsoons, and their consequences, to be mapped to keywords. Citizen scientists can also support in the evaluation of extracted images with crowdsourcing, as it has been proposed in the Crowd4EMS project (\cite{ravi2019crowd4ems}), as well as of the resulting keywords of the TriggerCit methodology (see Step~\ref{step:promising}). A software library could then be used to automate the aggregation of annotations (e.g., \cite{crowdnalysis2022}).

As discussed in the paper, Twitter has different usage rates in different countries. Alternatives to Twitter 
        include other social media 
        like Facebook, Telegram, and WhatsApp. These are used to facilitate communication between volunteers and emergency services, enable sharing of information within a community, providing swift updates on emergency situations in real-time. However, even though such platforms contain a wealth of information, access to data is usually restricted (e.g., for privacy concerns). On the other hand, systems like Facebook Disaster Maps provide aggregated insights for crisis response and recovery and its use should be further investigated. In addition, as some tweets contain URLs pointing to other contents, such as web pages (e.g., newspapers) or posts in other social networks,  these contents can also be crawled in order to obtain more information about the event. Moreover, the information extracted from one social media could be leveraged to crawl other platforms iteratively, as in \cite{Autelitano2019}. 
        

Further directions to be explored in order to enhance TriggerCit capabilities include:

\begin{itemize}     
        \item Training a supervised learning classifier for triggering \newline
        We are still in the phase of collecting high-quality time series data related to a representative number of flood events. Once a large dataset is available, we plan to experiment with supervised learning approaches, in order to automate the triggering decision and validate it against a variety of events (e.g., different scales and areas).
        Such a classifier should also include contextual data about the event, information about flooded areas (such as the ones automatically classified in \cite{FloodAI}), and about the readiness of the areas in terms of preparation measures for recurring flood events. 
        
        \item Automating dictionary management \newline
        Our system needs high-quality dictionaries for each language. Part of the future work will focus on automatically deriving event-specific keywords, both in general and adapted to a particular event. This automation would enhance the sensing phase, in which the system is monitoring social media for potential triggers. On the other hand, online adaptation to a specific event could increase the recall of the process.
        
    \end{itemize}

\section{Concluding Remarks}
The paper proposed a multilingual methodology to provide early flood alerts with associated images and locations. The results in two case studies in Asia in 2021 show that the alert is consistent in the Thailand case and that a timely evidence could have been provided for the Nepal floods.
Based on this analysis, TriggerCit could be deployed at UNOSAT into a cloud centralised service at CERN built on Kubeflow, a machine learning platform running on Kubernetes, allowing the integration with automatic flood detection from satellite data.

\paragraph{\bf Acknowledgements}
The work at Politecnico di Milano, University of Geneva and IIIA-CSIC was funded by the European Commission H2020 Project  Crowd4SDG,  
\#872944. 

The work at the United Nations Satellite Centre (UNOSAT) is part of the operational mapping service funded by the Norwegian Ministry of Foreign Affairs.

\printbibliography


\end{document}